\documentclass{article}

\usepackage{PRIMEarxiv}

\usepackage[utf8]{inputenc} 
\usepackage[T1]{fontenc}    
\usepackage{hyperref}       
\usepackage{url}            
\usepackage{booktabs}       
\usepackage{amsfonts}       
\usepackage{nicefrac}       
\usepackage{microtype}      
\usepackage{lipsum}
\usepackage{fancyhdr}       
\usepackage{graphicx}       
\usepackage{lineno}
\usepackage{amsmath}
\usepackage{graphicx}
\usepackage{rotating}
\usepackage{hyperref}
\usepackage{fancybox}
\usepackage{tikz}
\usepackage{upgreek}
\usepackage{makecell}
\usepackage{multirow}
\usepackage{xr}

\newcommand{\vecr}{\mathbf{r}}
\newcommand{\vecdr}{\delta\mathbf{r}}
\newcommand{\vecd}{\mathbf{d}}
\newcommand{\vecx}{\mathbf{x}}
\newcommand{\dvecx}{\Delta\mathbf{x}}
\newcommand{\vecb}{\mathbf{b}}
\newcommand{\mata}{\mathbf{A}}
\newcommand{\mats}{\mathbf{S}}
\newcommand{\matm}{\mathbf{M}}
\newcommand{\numpts}{n_p}
\newcommand{\numoffsets}{n_d}
\newcommand{\numnns}{n_{\text{NN}}}
\newcommand{\tike}{\emph{Tike}}
\newcommand{\ptychoshelves}{\emph{PtychoShelves}}
\newcommand{\ptychonn}{\emph{PtychoNN}}

\newcommand{\micron}{$\upmu$m}

\DeclareMathOperator*{\argmin}{arg\,min}

\graphicspath{{figures/}}

\title{Predicting ptychography probe positions using single-shot phase retrieval neural network
}

\author{
  Ming Du \\
  Advanced Photon Source\\
  Argonne National Laboratory\\
  Lemont, IL 60439, USA \\
  \texttt{mingdu@anl.gov}
   \And
  Tao Zhou \\
  Center for Nanoscale Materials\\
  Argonne National Laboratory\\
  Lemont, IL 60439, USA \\
  \texttt{tzhou@anl.gov}
   \And
  Junjing Deng, Daniel J. Ching, Steven Henke \\
  Advanced Photon Source\\
  Argonne National Laboratory\\
  Lemont, IL 60439, USA \\
 \And
  Mathew J. Cherukara \\
  Advanced Photon Source\\
  Argonne National Laboratory\\
  Lemont, IL 60439, USA \\
  \texttt{mcherukara@anl.gov}
}

\begin{document}

\textbf{GOVERNMENT LICENSE}

The submitted manuscript has been created by UChicago Argonne, LLC, Operator of Argonne
National Laboratory (``Argonne''). Argonne, a U.S. Department of Energy Office of Science laboratory, is operated under Contract No. DE-AC02-06CH11357. The U.S. Government retains for
itself, and others acting on its behalf, a paid-up nonexclusive, irrevocable worldwide license in
said article to reproduce, prepare derivative works, distribute copies to the public, and perform
publicly and display publicly, by or on behalf of the Government. The Department of Energy will
provide public access to these results of federally sponsored research in accordance with the DOE
Public Access Plan. http://energy.gov/downloads/doe-public-access-plan.
\newpage

\maketitle

\begin{abstract}
Ptychography is a powerful imaging technique that is used in a variety of fields, including materials science, biology, and nanotechnology. However, the accuracy of the reconstructed ptychography image is highly dependent on the accuracy of the recorded probe positions which often contain errors. 
These errors are typically corrected jointly with phase retrieval through numerical optimization approaches. When the error accumulates along the scan path or when the error magnitude is large, these approaches may not converge with satisfactory result. We propose a fundamentally new approach for ptychography probe position
prediction for data with large position errors, where a neural network is used to make single-shot phase retrieval on individual diffraction patterns, yielding the object image at each scan point. The pairwise offsets among these images are then found using a robust image registration method, and the results are combined to yield the complete scan path by constructing and solving a linear equation. We show that our method can achieve good position prediction accuracy for data with large and accumulating errors on the order of $10^2$ pixels, a magnitude that often makes optimization-based algorithms fail to converge. For ptychography instruments without sophisticated position control equipment such as interferometers, our method is of significant practical potential. 
\end{abstract}


\section{Introduction}

Ptychography is a coherent diffractive imaging technique that uses the diffraction patterns of a specimen to reconstruct its image. The technique involves scanning a specimen with a probe, typically a beam of light or electrons, and recording the resulting diffraction patterns. These patterns are then used to reconstruct the image of the specimen. As a scanning microscopy technique, ptychography relies on accurately recorded probe positions to achieve high-quality reconstructions. The probe positions are typically determined before the
experiment by taking points from a certain scan trajectory, \emph{e.g.}, a raster grid, 
a concentric spiral, or a Fermat spiral pattern \cite{huang_opex_2014}. The points are visited
sequentially during the imaging experiment, and a diffraction pattern is collected at each point. 
In reality, the positions where the diffraction patterns are actually measured might be different from
the expected coordinates due to mechanical or thermal perturbation to the imaging system, 
or the imprecision of the sample stage actuator. These errors are one of the contributing factors
to degraded reconstruction quality. 

Depending on the source of error and the capability of experimental control devices, ptychography
probe positions can be divided into two categories. They are:
\begin{itemize}
    \item Accumulating errors, where errors in points acquired earlier in the scan are passed over
    to later points. Denoting the accumulating position error at point $i$ as $\vecdr_i$, 
    then $\vecdr_i$ is
    contributed by $\sum_{j \leq i}\vecdr_i$. This type of error can emerge if the sample stage
    actuator is not equipped with an encoder and thus runs without closed-loop positional control. 
    \item Independent errors, where the error occurring at one scan point does not affect other
    scan points. This type of error is typically random and can be contributed by mechanical
    perturbations or imprecision of sample stage motor readout. 
\end{itemize}

Based on the Fourier shift theorem, translations of a real-space signal are encoded in the phase of
its reciprocal space representation. Since only intensities are recorded in the diffraction patterns, calculating the precise relative translational relation between diffraction patterns
cannot be done without phase retrieval. 

Many existing methods used for probe position correction
refine the positions simultaneously with phase retrieval, where the positions are updated before
or after each iteration of update for the object function or probe function. One way of determining
the update equation for the positions is to treat phase retrieval and position correction
as a joint non-linear optimization problem, and update the positions using
gradient-based optimization methods like steepest gradient descent or conjugate gradient descent
\cite{guizarsicairos_opex_2008, odstrcil_opex_2018, tripathy_opex_2014}. Another method described in 
\cite{dwivedi_ultramicroscopy_2018} approximates the error of diffraction pattern intensity
(defined as the difference between the measured intensity and the predicted intensity at the
current iteration) to be linearly related to probe position errors, and solves the position errors at each
iteration of phase retrieval. Derivative-free methods based on genetic algorithm \cite{shenfield_jap_2011},
and simulated annealing \cite{maiden_ultramic_2012} have also been proposed. Additionally, 
\cite{beckers_ultramicroscopy_2013} describes a drift model fitting method applicable to situations where
the position errors are systematic and parameterizable. 

Methods based on non-linear or linearly approximated optimization are widely accepted in the community
and are implemented in several well-established ptychography libraries such as PtychoShelves 
\cite{wakonig_jac_2020} and PyNX \cite{favrenicolin_jac_2020}. However, we have observed that
these methods either fail or take long to converge at a good solution 
when the error magnitude is large, or when the error is accumulative. To understand the failure
mechanism, one may imagine a situation where the position error is so large that two diffraction patterns
that are thought to be close to each other in fact have little overlap. This breaks the information
redundancy constraint which ptychography relies on to recover the phase of the object and the probe.
Then, if one considers from the perspective of gradient-based optimization, the gradient to probe
position $\vecr_i$ can be expressed as the spatial gradient of the object 
$\partial o(\vecr - \vecr_i) / \partial \vecr_i$ multiplied by the gradient to the object
$\partial L / \partial o(\vecr - \vecr_i)$, where $L$ is the reconstruction loss function. If
$\partial L / \partial o(\vecr - \vecr_i)$ is not reliable enough due to the loss of information
redundancy constraint, then the gradient for the probe position would be unreliable too. 
The derivative-free methods mentioned above could potentially avoid this pitfall, but considering
the high dimensionality of probe positions' solution space, they are computationally very
expensive for large scan grids, and are so far demonstrated only for relatively small number of
scan points. 

If the real-space image at each scan point is available, then one can exploit commonly available
image registration algorithms to find out the positional relation between pairs of scan points, 
which can be used to correct, or even predict from scratch the probe positions. This philosophy has
in fact been explored in several works. In \cite{zhang_opex_2013}, for each scan point, the 
local object function is registered with the object in the previous iteration to estimate the
position error for that scan point, based on the observation that the local object function can
shift to its true position by itself when the position error is small. However, due to the
small-error assumption, this method may not work when the error is on the level of several tens of
pixels. In \cite{rong_opex_2019}
and \cite{dwivedi_jopt_2019}, a more straightforward approach is used, where at each phase 
retrieval iteration, the local object function at each scan point is registered with its neighbor
to update the position of that point. There are several concerns about this approach: first, 
since the images used for registration are collected during phase retrieval, the low image quality
at the early stage might make registration difficult. Second, since the object function is only
effectively updated in the region strongly illuminated by the probe, a large amount of probe overlap
is required for image registration to be accurate. When the position errors are too large to
maintain a decent overlap between certain pairs of probes, this method may fail. Therefore, 
correcting ptychography probe positions with large or accumulating errors, where the distance
between actual positions and expected positions can be as large as several tens or hundreds of
pixels, remains an open question. 

Deep learning has opened a new portal to this problem. In the coherent imaging community, deep
learning-based algorithms have been shown to be able to produce high-quality phase retrieval
results given measured diffraction intensities in a non-iterative fashion, being demonstrated
for inline holography \cite{rivenson_lsa_2018}, Bragg coherent diffraction imaging \cite{yao_npjcompmat_2022, chan2021rapid}, 
and ptychography \cite{cherukara_apl_2020, babu2023deep}. The \ptychonn{} algorithm 
described in \cite{cherukara_apl_2020} has proven capable of generating accurate reconstructions
with individual diffraction patterns (\emph{i.e.}, diffraction patterns collected with overlapping
probes are not needed) after training.  
This unique capability, referred to as ``single-shot'' phase retrieval,
relaxes the traditional requirement of ptychography that phase retrieval is only possible with
sufficient probe overlap, and is a result of the network's implicit learning of the probe
function and the object function's distribution, which constitute the prior knowledge that
compensates the lack of probe overlap. 

\begin{figure}
    \centering
    \includegraphics[width=0.9\textwidth]{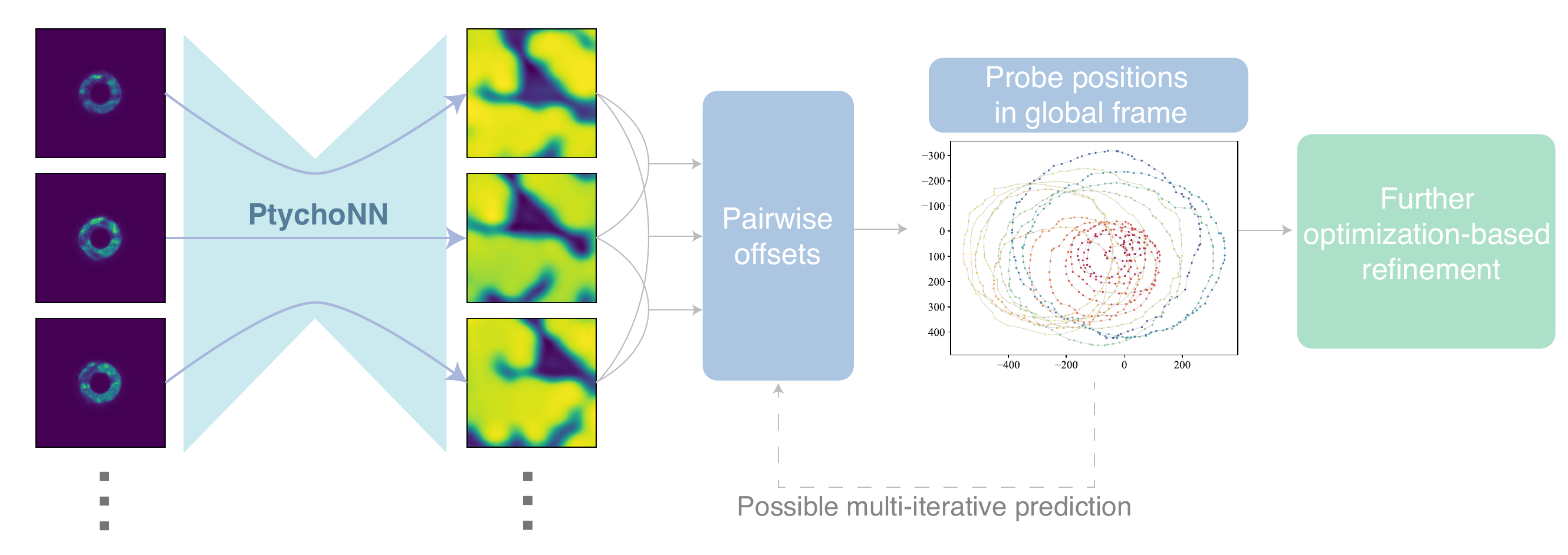}
    \caption{Workflow of our proposed probe position prediction method based on single-shot
    phase retrival and pairwise image registration.}
    \label{fig:workflow}
\end{figure}

With the single-shot phase retrieval capability of \ptychonn{}, we can now build a fundamentally new
approach to the probe position correction problem that contains two stages. 
In the first stage, given a trained \ptychonn{} model, we do
single-shot predictions on each individual diffraction pattern of the test data. 
In the second stage, we use
image registration algorithms to find out their pairwise spatial offsets. With this, we can
solve the positions of all scan points in the same global frame. This also means we can
reconstruct the entire scan trajectory completely from scratch, without any probe position information 
as input. Therefore, the method we propose not only ``corrects'' the probe positions, but
also ``predicts'' them. The predicted probe positions can be further refined using optimization-based
algorithms during phase retrieval.
The workflow of our approach is summarized in Fig.~\ref{fig:workflow}. 

In the following sections, we describe how our method can predict the probe positions in cases
with either accumulating errors or independent errors. The average pairwise offset error
between the predicted positions and their corresponding true positions is on the order of a few pixels,
and this is true even in the most extreme cases of accumulating errors, where the distance of the
errors is on the order of $10^1$ -- $10^2$ pixels. This level of error typically precludes optimization-based
algorithm from converging, but if instead they start with the positions predicted
by our method, the positions can be further refined and the average pairwise error could be reduced
to single or sub-pixel level. As such, our method can and is intended to work synergistically
with existing optimization-based position correction algorithms. 

\section{Methods and Materials}

\subsection{Data acquisition and processing}

\subsubsection{Experimental setup}

Two datasets collected from different samples were used in this work. Both of them were collected at the Advanced Photon Source (APS). For the first dataset, the sample was a custom designed calibration chart containing random etched patterns fabricated in 1.5 \micron{} tungsten by Zoneplates Ltd. This sample will be hereafter referred to as the
``random etched object''. The data was collected at the
hard x-ray nanoprobe at the APS with a photon energy of 10 keV. A Fresnel Zone Plate was capable of focusing the x-ray beam to 25 nm. The sample was intentionally placed at 120 \micron{} downstream of the focal point to achieve a larger beam size of about 800 nm FWHM. Ptychographic scans were performed following a spiral trajectory. For the training data, 12 isolated regions were probed, each with three spiral scans of  different step sizes, 50 nm, 100 nm and 200 nm. The number of diffraction pattern per spiral scan, collected on a Medipix 3 detector at 0.9 m from the sample, was 961. The maximum number of training data was thus 34596. The test dataset was collected in a region well separated from where the training data was collected. The step size of the spiral scans for the test dataset also varies from 50 nm, to 100 and 200 nm. Additionally, random positional error was intentionally injected in the experimental scan trajectory. The positional error follows a Gaussian distribution with a standard deviation corresponding to 0 (no error), 0.1, 0.2, 0.4, and 0.8 times the corresponding step size.
In this way, the scans of the test data contain position errors if one were to assume they were collected using the ideal spiral scan, but the errors are known by design, and the actual scan positions are precisely controlled by
a laser interferometer, ensuring that the designed scan positions can be used as ground truths.
The table below lists the conditions for each test dataset. 

\begin{table}[]
    \centering
    \begin{tabular}{cccc}
        \hline
        \textbf{Label} & \textbf{Error type} & \textbf{Nominal step size (nm)} & \makecell{\textbf{Error standard deviation}\\ \textbf{(multiples of step size)}} \\
        \hline
        A1-1 & Accumulating & 200 & 0.1 \\ 
        A1-2 & Accumulating & 200 & 0.2 \\ 
        A1-3 & Accumulating & 200 & 0.4 \\ 
        A2-1 & Accumulating & 100 & 0.1 \\ 
        A2-2 & Accumulating & 100 & 0.2 \\ 
        A3-1 & Accumulating & 50 & 0.1 \\ 
        A3-2 & Accumulating & 50 & 0.2 \\ 
        A3-3 & Accumulating & 50 & 0.4 \\ 
        I1-1 & Independent & 200 & 0.1 \\ 
        I1-2 & Independent & 200 & 0.2 \\ 
        I1-3 & Independent & 200 & 0.4 \\ 
        I1-4 & Independent & 200 & 0.8 \\ 
        I2-1 & Independent & 100 & 0.4 \\ 
        I2-2 & Independent & 100 & 0.8 \\ 
        I3-1 & Independent & 50 & 0.4 \\ 
        I3-2 & Independent & 50 & 0.8 \\ 
        \hline
        
    \end{tabular}
    \caption{Parameters and error conditions used for each test dataset.}
    \label{tab:test_sets}
\end{table}

The second test sample is a Siemens star test pattern of 500 nm thick gold imaged using
a scanning x-ray microscope at beamline 2-ID-D at the APS. Details on the data acquisition
experiment was also documented in \cite{du_opex_2021}, and readers are recommended to refer to
Section 3.4 of the published paper. Notably, the data were collected 
with a step size of 50 nm and 
a scan grid of $70\times 70$. Without the use of advanced stage positioning
instruments, the actual positions involve random errors that are unknown. 

\subsubsection{Data curation}

As mentioned above, our training set contains 36 scans and 
each scan contains 961 scan points, giving a total of 34596 diffraction patterns. 10\% of these data were split out as the validation set.
The original $512\times 512$-sized diffraction patterns 
were cropped to $384\times 384$ and then downsampled by 3 times, resulting in a final size of 
$128\times 128$. To generate training labels, we reconstructed all the scans in the training set
using the rPIE \cite{maiden_optica_2017} algorithm in \emph{Ptycholib},
a ptychographic reconstruction library \cite{nashed_optexp_2014}. 
Labels were then obtained by extracting a $128\times 128$-sized region from the reconstructed
phase through bilinear interpolation for each scan point,
assuming a pixel size of 8 nm. 

For the Siemens star object, we took all the even rows in the $70\times 70$ scan grid as
the training set, resulting in 2450 diffraction patterns. Since this dataset has been reconstructed
with gradient-based probe position correction for the demonstration of 
\emph{Adorym} \cite{du_opex_2021}, 
an inverse problem solving framework based on automatic differentiation, we generated the labels
by extracting $128\times 128$-sized regions from the phase image reconstructed by \emph{Adorym}
according to the final positions refined by it. 
The original reconstruction has a pixel size of 13.2 nm,
and we resampled it with a new pixel size of 16 nm
when generating the labels.
For the test set, we took all the odd rows
between row 9 and row 45, where the local phase images are more diverse in the directions of edges
and are thus more suitable for image registration. 

Feature engineering has also been conducted to the diffraction patterns for both samples. We took
the mean over the entire training set for each pixel in the reciprocal domain, obtaining a
``mean diffraction pattern'', and subtract it from every diffraction pattern. In this way, we
not only eliminate the common components among the diffraction patterns, allowing the network
to focus on the variation across scan points, but also reduce the dynamic range of data, improving
training stability. After that, we standardized the diffraction patterns using the mean and 
standard deviation across the entire training set. 

\subsection{\ptychonn{}}

\subsubsection{Architecture design}

The architecture of \ptychonn{} used in this work is primarily adopted from the earlier publication
\cite{cherukara_apl_2020}. However, several modifications and improvements were implemented in
our work. The vanilla \ptychonn{}
described in the cited paper employs an autoencoder-like architecture with 2 decoders sharing the 
same structure respectively used
for reconstructing the phase and magnitude of the object function. In the original work, the sizes
of the input diffraction patterns and the output phase and magnitude images are $64\times 64$,
and both the encoder and decoders involves 3 $2\times 2$ max-pooling downsampling operations. To work on 
$128\times 128$-sized data in our work, we increased the number of scale levels by 1, using 4
max-pooling layers. Also, since the quality of phase prediction is generally better than magnitude with hard x-rays,
we removed the magnitude encoder, making \ptychonn{} predict only the phase. Additionally, we added
batch normalization \cite{ioffe_arxiv_2015} after each convolution layer to stabilize training.

\subsubsection{Training and prediction}

The network was trained on 2 NVIDIA Quadro P4000 GPUs. The training was set to run for 60 epochs, 
using a batch size of 64 per GPU. L1 loss measured between predicted and true phases is used as
the objective. 
Adam \cite{kingma_arxiv_2014} was used as the 
optimizer with an initial learning rate of $10^{-4}$. The learning rate was scheduled to 
change following a cyclic and decaying 
triangular function (\emph{triangular2} in \cite{smith_arxiv_2015}) 
throughout the training process, with a base learning rate of $10^{-5}$ and a cycle length of 
12 epochs. The chosen batch size and initial learning rate were determined through a grid search of 
hyperparameters. The model after the epoch with the lowest validation loss was chosen to be used
for prediction on the test datasets. 

\subsection{Image registration}

An image registration algorithm that performs accurately and robustly on phase images predicted by
\ptychonn{} is the key for our method to succeed. To fulfill this requirement, we constructed a hybrid
registration algorithm that is consisted of an ergodic algorithm named ``error map'' and the well-known
Scale-Invariant Feature Transform (SIFT) algorithm \cite{lowe_procieeecv_1999}.

\subsubsection{Error map}

Denoting the moving and reference image as $I_m(x, y)$ and $I_r(x, y)$, the error map algorithm rolls 
$I_m(x, y)$ by
every possible integer shift vector $(d_x, d_y)$ within $d_{x,\min} \leq d_x \leq d_{x,\max}$
and $d_{y,\min} \leq d_y \leq d_{y,\max}$. Rolling an image with a finite support would result in pixels
wrapping around from the other side. However, 
for an image with support size $(N_x, N_y)$,
as long as $d_{x,\max} - d_{x,\min} < N_x$ and 
$d_{y,\max} - d_{y,\min} < N_y$, there is always a region of pixels $(x, y) \in \Omega$ that is never
affected by wrapping. For each shift vector, we calculate the mean squared error (MSE) between
the rolled image $I_m(x-d_x, y-d_y)$ and reference image $I_r(x, y)$ within the unwrapped region.
The integer offset between the 2 images is then estimated as

\begin{equation}
    \vecd_{\text{int}} = \argmin_{(d_x, d_y)} \sum_{(x,y)\in\Omega}\left[ I_m(x - d_x, y - d_y) - I_r(x, y) \right]^2.
\end{equation}

Along with calculating the minimal MSE, we also construct an error map $E(d_x, d_y)$ where every pixel
stores the MSE for a certain shift vector. To obtain sub-pixel precision, we fit a 2D quadratic function
around the minimum of $E(d_x, d_y)$. The precise offset can be then derived from the parameters of the 
quadratic function. 

We note that compared to Fourier signal analysis-based registration methods like phase correlation 
\cite{reddy_ieeetransimageproc_1996}, the ``error map'' method excludes the wrap-around regions in metric
calculation, and is thus more reliable for small-sized images.

\subsubsection{SIFT}

The SIFT algorithm is documented in \cite{lowe_procieeecv_1999}. SIFT uses a descriptor-based
method to find multiple keypoints in both images to be registered, matches keypoints with similar
descriptors, and find out the transformation between the coordinates of the two sets of matched keypoints.
Since we are only interested in the translational offset between the images, we simply calculate the
averaged difference between the coordinates of all pairs of matched keypoints. 
The over-the-shelf SIFT algorithm, however, often produces
erroneously matched key points which skew the registration results on the images predicted by
\ptychonn{}. Therefore, we filter the matched keypoints by applying the offset given by each pair of
the points, and reject the match if the MSE within the unwrapped region between the shifted and reference
images is above a certain threshold. 

\subsubsection{Hybrid registration}

Based on our observation, the error map algorithm is more accurate for image pairs with small offsets.
When the offset is large, however, the result becomes less reliable and less stable because the unwrapped
region $\Omega$ becomes so small that it is vulnerable to artifacts, local distortions, and low-variance
(featureless) regions. Our hybrid algorithm runs error map as the first pass trial, and apply the
calculated offset to $I_m$ and compute the error within the unwrapped region. If the error is larger
than a certain threshold or if the variance in the unwrapped region is below a certain threshold
(indicating the region is too ``flat'' to yield a unique solution), we register the images using SIFT
instead, which would typically give a less accurate but robust solution. 

\subsection{Solving probe positions}

Image registration only gives pairwise offsets between images collected from scan points close to each
other. We introduce the following two modes in which our algorithm operates to solve for the positions 
of all scan points in the global frame.

\subsubsection{Serial mode}

In serial mode position calculation, we assume the global position of the first scan point (point 0) 
is (0, 0). Then starting from point 1, we register the image at point $i$ with the image at point
$i - 1$, obtaining the offset vector $\vecd_{i/i - 1}$. The position of point $i$ is then given by
$\vecr_{i} = \vecr_{i - 1} + \vecd_{i/i - 1}$. In this mode, the position of each scan point solely
depends on the previous point. 

Occasionally, when the offset between a pair of consecutive scan points is very large, or when the images
are too inconsistent or lacking of features, the registration result could be unreliable even with our
hybrid registration method. 
We found that when this happens, it is better to use a good guess as
the current offset than to use the registration result directly. We introduce here a momentum-based
offset estimator, inspired by momentum-based numerical optimization 
algorithms \cite{qian_neuralnetworks_1999}, for generating offset guesses. Throughout the
serial mode calculation process, we keep track of a running average of the offset, $\overline{\vecd}$. 
At point $i$, $\overline{\vecd}$ is updated as
\begin{equation}
    \overline{\vecd} \leftarrow \beta\overline{\vecd} + (1 - \beta)\vecd_{i/i - 1}.
\end{equation}
The reliability of the registration result can be characterized by
the error in the unwrapped region after applying the offset to the moving image. If this error is too large, we use the estimated offset $\overline{\vecd}$ at that point. It is
possible to build estimate algorithms using higher-order momenta, but in practice, we found the first-order
algorithm generally yields the best results. 

The biggest drawback of the serial mode is that the error accumulates along the scan
trajectory. Especially due to the fact that the position of a point entirely depends on the previous
point, the error in any registration result inevitably influences all the subsequent points. Therefore,
the serial mode is only used to obtain a preliminary set of positions when the raw motor readout
is unavailable or unreliable. To improve the accuracy of the result, we propose the algorithm
introduced below, named as ``collective mode''. 

\subsubsection{Collective mode}

Assuming the probe positions recorded by the motor is available and not too erroneous, or a serial mode
position calculation already has generated a preliminary set of probe positions, for each scan 
point $i$, we can find out several points located closest to it. Together with the 2 points measured
right before and after point $i$ in the scan trajectory, we denote this set of $k$ points as the 
nearest neighbors of point $i$. The image at point $i$ is registered with each of its neighbors ($j$),
provided that $i$ and $j$ have not been registered before. Theoretically, this would yield a total of
$k\numpts / 2$ registration results, where $\numpts$ is
the total number of scan points. However, we should reject the registration results that are 
unreliable (detected by the error thresholding method mentioned above), so the number of valid
registration result $\numoffsets$ would be generally smaller than that. Denoting $\vecr$ to be a
$\numpts \times 2$ matrix containing the positions of all scan points ($\vecr_i$), $\vecb$ to be a 
$\numoffsets \times 2$ matrix containing the all the offsets found, then the probe positions can be
found by solving the following equation:
\begin{equation}
    \mata\vecx = \vecb
\end{equation}
where $\mata$ is a $\numoffsets \times \numpts$ matrix with each row $m$ corresponding to an offset
$\vecd_{i/j} = \vecr_j - \vecr_i$, and
\begin{equation}
    A_{mn} = 
    \begin{cases}
        1,\ n = j \\ 
        -1,\ n = i \\
        0,\ \text{otherwise}
    \end{cases}.
\end{equation}

The linear equation above needs to be solved with proper constraints. Since the collective mode
has an initial set of probe positions $\vecx_0$ as its input (for determining nearest neighbors), 
we may solve for the residue between $\vecx$ and $\vecx_0$, $\dvecx$, instead. With two additional regularization terms, we find $\dvecx$ by solving the
following least-squares problem:
\begin{equation}
    \widehat{\dvecx} = \argmin_{\dvecx} \|\mata(\vecx_0 + \dvecx) - \vecb\|^2 + \lambda_1\|\dvecx\|^2 + 
    \lambda_2\|(\vecx_0 + \dvecx) - \mats(\vecx_0 + \dvecx)\|^2
    \label{eqn:lsq}
\end{equation}
where $\lambda_1$ and $\lambda_2$ are weights of the regularization terms;  
$\mats$ is a shift matrix, and $\mats\vecx$ rolls $\vecx$ up by 1 element. 
Assuming $\vecx_0$ is 
a reasonably good estimate of the probe positions, the first regularization term penalizes the
2-norm of $\dvecx$. The second regularization term is a smoothness constraint that penalizes the
difference between the 2 adjacent probe positions. Usually, $\lambda_1$ is kept to a small value
($10^{-6}$) and $\lambda_2$ is set to 0 to avoid overconstraining the solution, but they may be increased
in the event that the solution is unstable. The solution to this least-squares problem is given by
\begin{equation}
    \widehat{\dvecx} = (\mata^T\mata + \lambda_1\mathbf{1} + \lambda_2\matm)^{-1}(\mata^T \vecb - \mata^T\mata\vecx_0 - \lambda_2\matm\vecx_0)
\end{equation}
where $\matm = (\mathbf{1} - \mats)^T(\mathbf{1} - \mats)$. 

\subsubsection{Multi-iteration prediction}

The serial and collective mode of position prediction can be run for multiple times sequentially, with the solution of the previous iteration used for nearest neighbor search and as
the $\vecx_0$ in Eq.~\ref{eqn:lsq}.

As a general guideline, if the nominal positions determined
before the experiment or the positions from the motor readout are unreliable, one may start with 
a serial mode prediction to estimate the scan points' relative locations, then following that with
1 -- 2 iterations of collective mode prediction to refine the positions. Otherwise, one may choose
to skip serial mode and directly run collective mode using nominal positions as the input
for neighbor search. 

\subsection{Ptychography reconstrution and further position refinement}

To evaluate the results of position prediction, we performed ptychography reconstructions on the test
data listed in Table \ref{tab:test_sets}. The reconstructions for our primary assessment are done
using \tike \cite{tike}, a ptychography data processing framework developed at the Advanced Photon Source. \tike{} provides multiple reconstruction algorithms including rPIE
\cite{maiden_optica_2017}, which is the algorithm we use for our data. \tike{} also offers the
option of refining probe position using gradient-based and momentum-enabled Adam \cite{kingma_arxiv_2014}
algorithm. For each test dataset, we performed the following reconstructions:
\begin{enumerate}
    \item With true probe positions, without further position refinement;
    \item With nominal positions, without further position refinement;
    \item With nominal positions, with further position refinement during reconstruction;
    \item With predicted positions, without further position refinement;
    \item With predicted positions, with further position refinement during reconstruction.
\end{enumerate}
To verify the results obtained using \tike{}, we also conducted reconstructions for nominal and predicted
positions with further position refinement using \ptychoshelves \cite{wakonig_jac_2020}, where position
refinement is implemented following \cite{odstrcil_opex_2018}. The results were cross-validated
with the results obtained using \tike{} and
are shown in Fig.~S1 and S2 in Supplemental Materials. 

\section{Results}

\subsection{Random etched object}
\label{sec:random_etch_sample}

\begin{figure}
    \centering
    \includegraphics[width=0.8\textwidth]{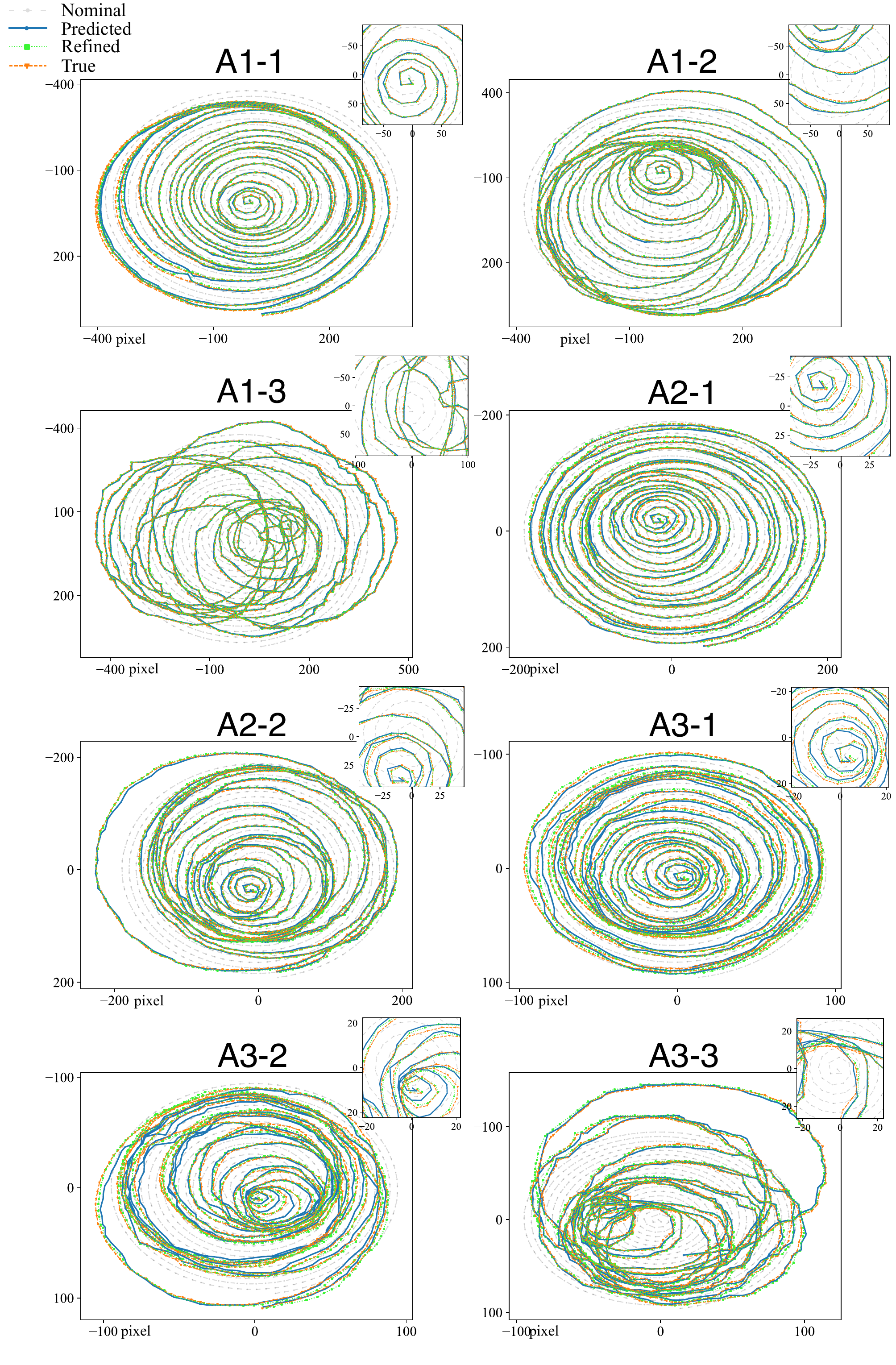}
    \caption{Comparison of the nominal positions, true positions, predicted positions, and
             positions after refinement initialized with predicted positions for
             all cases with accumulating errors in 
             Table \ref{tab:test_sets}. On the top right of each subplot is
             the magnified sub-region around the graph's center. 
    }
    \label{fig:random_etch_paths_accumulating}
\end{figure}

\begin{figure}
    \centering
    \includegraphics[width=0.8\textwidth]{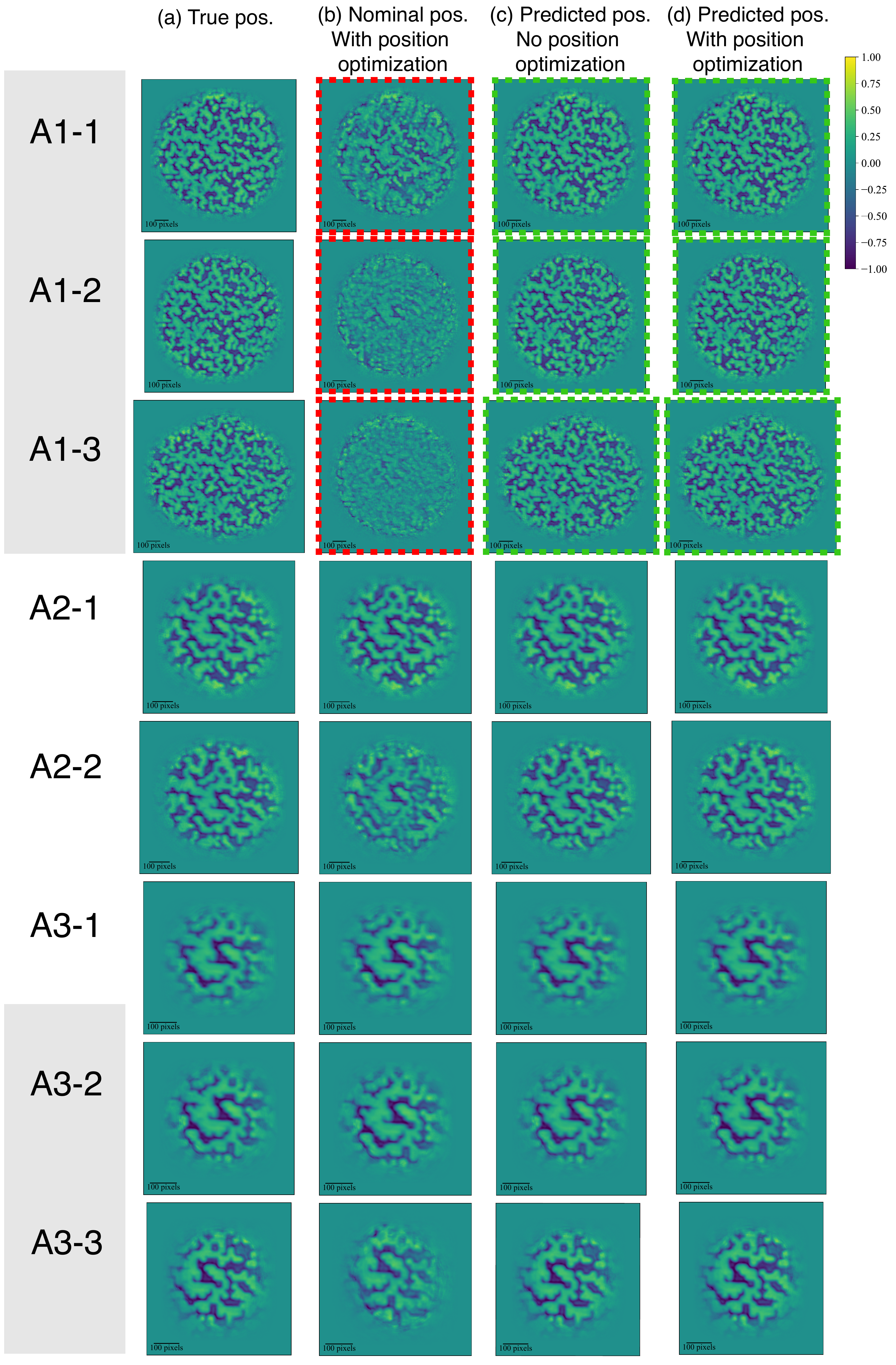}
    \caption{Ptychography reconstruction results of all cases with
             accumulating errors in 
             Table \ref{tab:test_sets}. 
             Column (a) -- (d) shows the reconstructed phase image using true, nominal, and predicted
             positions with or without further refinement using optimization-based algorithms. Case A1-1 -- A1-3 are
             most representative for cases with large accumulating errors as demonstrated by the sharp contrast between
             reconstructions using nominal positions (red dash-boxed images) and predicted positions (green dash-boxed images).
    }
    \label{fig:random_etch_recons_accumulating}
\end{figure}

\begin{figure}
    \centering
    \includegraphics[width=0.8\textwidth]{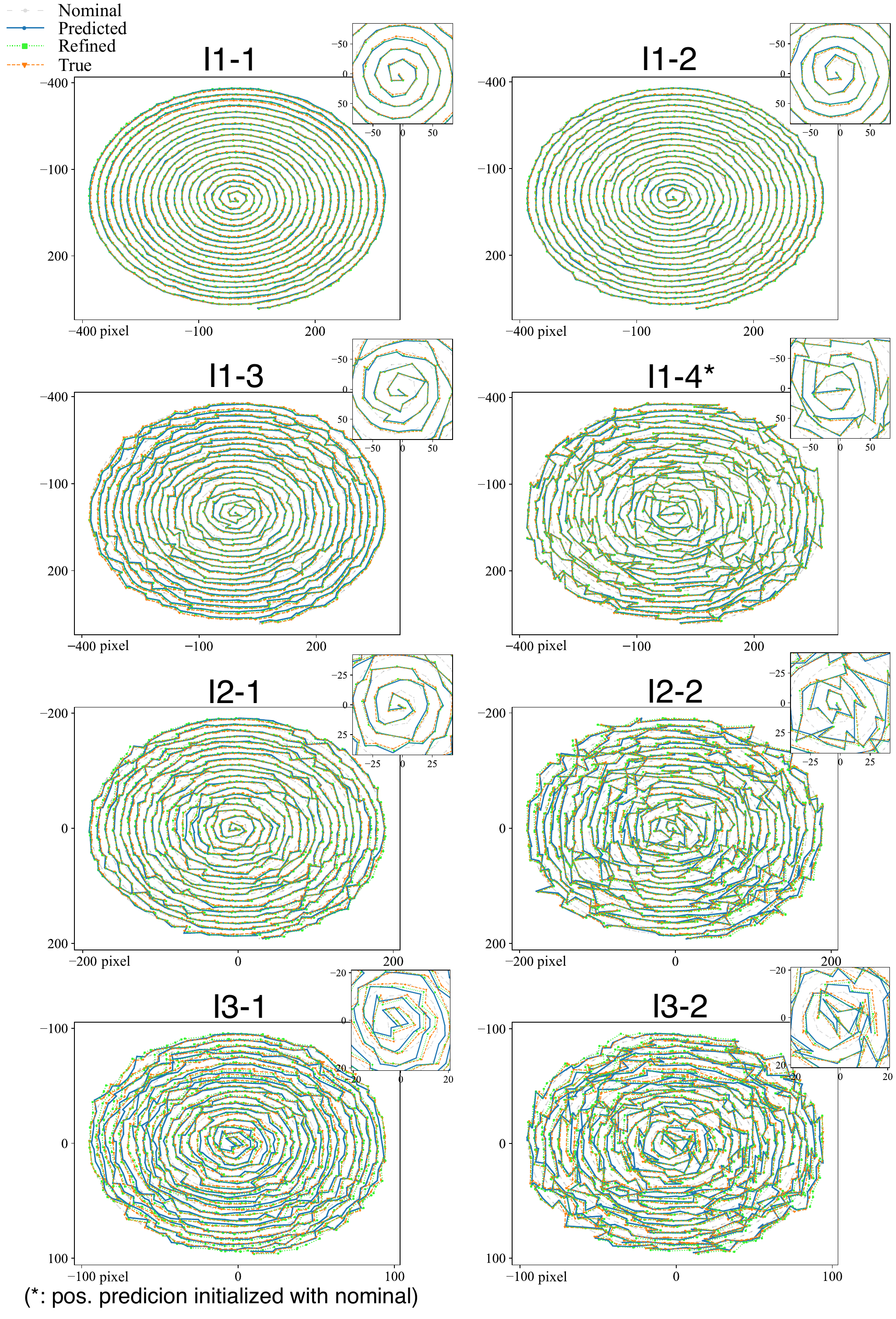}
    \caption{Comparison of the nominal positions, true positions, predicted positions, and
             positions after refinement initialized with predicted positions for
             all cases with independent errors in 
             Table \ref{tab:test_sets}. See the caption of Fig.~\ref{fig:random_etch_paths_accumulating} for description of the figure's elements.
    }
    \label{fig:random_etch_paths_independent}
\end{figure}

\begin{figure}
    \centering
    \includegraphics[width=0.8\textwidth]{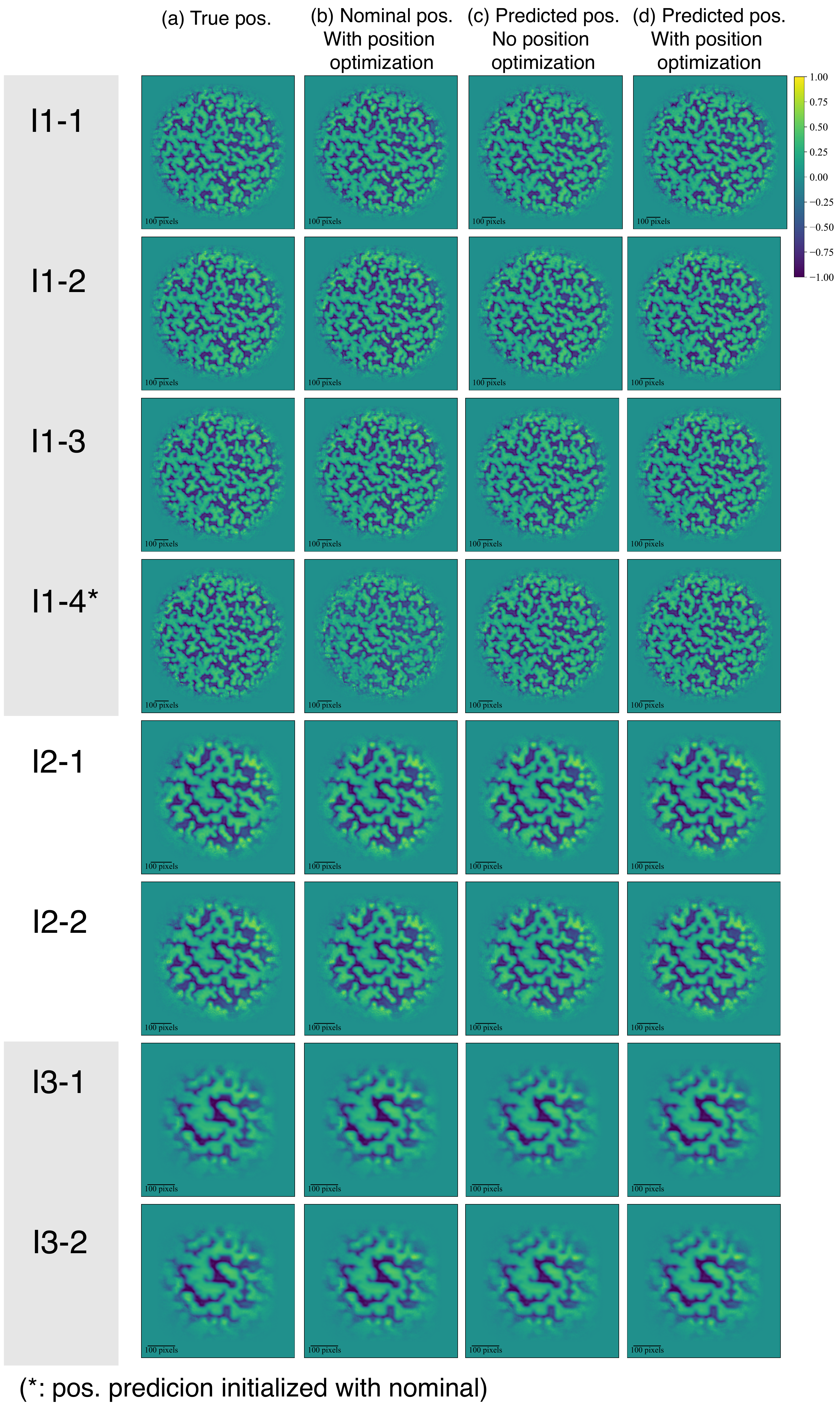}
    \caption{Ptychography reconstruction results of all datasets 
             with independent errors in Table \ref{tab:test_sets}. See the caption of 
             Fig.~\ref{fig:random_etch_recons_accumulating} for description of the figure's elements. 
    }
    \label{fig:random_etch_recons_independent}
\end{figure}

To quantitatively measure the accuracy of probe positions, we first introduce a 
metric that will be hereafter
referred to as the root-mean-square of pairwise position error (RMS-PPE). Since ptychography probe positions are relative, using an
absolute metric such as the mean squared error between two sets of positions can introduce arbitrary
bias even after subtracting the mean values from both sets. Additionally, if the position error
accumulates, the error in one scan point would be repeatedly counted in all the subsequent scan points.
To address these concerns, RMS-PPE is proposed as a metric measuring the error of relative, rather than absolute,
positions. 

RMS-PPE is calculated between an test and a true set of positions, with the assumption that the scan sequence of
all points are consistent across both sets. For each point in the true position set,
we find the indices of the $\numnns$ nearest neighbors, and calculate the offset of each of them to the current point. 
The offsets are calculated for the same pairs of indices in the test position set. This yields 2 sets of offsets,
and the root-mean-square (RMS) between them is calculated to give the RMS-PPE. In mathematical notations
this is written as
\begin{equation}
    \mbox{RMS-PPE} = \sqrt{\frac{1}{\numpts\numnns}\sum_{i=1}^{\numpts}\sum_{k\in \text{NN}_i} \left\| \vecd_{i/k,\text{test}} - \vecd_{i/k,\text{true}} \right\|^2}
\end{equation}
where $\text{NN}_i$ is the set of indices that are the $\numnns$-nearest neighbors of point $i$,
and $\vecd_{i/k}$ is the offset between point $i$ and
$k$ in a certain position set.
For the rest of this paper, we use $\numnns = 3$ when calculating RMS-PPEs, based on the
principle of triangulation positioning that the distances to at least 3 nearby known points
are needed to uniquely determine the position of a point. 

For all datasets collected from the random etched object
except I1-4, position prediction was run from scratch with 1 iteration of serial
mode and 2 iterations of collective mode, without using the nominal position at all. Serial mode
prediction is challenging for I1-4 due to the large pairwise offset magnitude, and since the error
does not accumulate, we used the nominal positions to initialize collective mode
for this only case. 

In Fig.~\ref{fig:random_etch_paths_accumulating} and \ref{fig:random_etch_paths_independent},
the nominal positions, true positions, the positions
predicted by our method, and the positions eventually refined by 
\tike{} with the predicted positions
used as the initial guess, are plotted for cases with accumulating
(Fig.~\ref{fig:random_etch_paths_accumulating}) and independent
(Fig.~\ref{fig:random_etch_paths_independent}) errors. It can be seen that
the positions predicted without any optimization-based refinement is are already close
to the true positions. Ptychographic reconstructions
initialized from true, nominal and predicted positions, with and without
position refinement, are shown in Fig.~\ref{fig:random_etch_recons_accumulating} and \ref{fig:random_etch_recons_independent}.
Using the predicted positions for ptychographic reconstruction, we
could already obtain sharp and clear images without doing further position refinement using
optimization-based methods, whereas for the 3 cases with large accumulating errors (A1-1 -- A1-3),
the reconstruction failed when initialized with 
nominal positions and
even with position refinement turned on (comparing images in Fig. \ref{fig:random_etch_recons_accumulating}(b) 
and (c) highlighted by dashed-line boxes). For most cases with 
small accumulating (A2-1 -- A3-3) and independent (I1-1 -- I3-2) errors, refinement could
find the right positions when initialized with nominal positions as it does with predicted
positions, but as we will see later, using predicted positions to initialize refinement still
leads to more favorable convergence behavior for almost all of these cases. 

\begin{figure}
    \centering
    \includegraphics[width=0.7\textwidth]{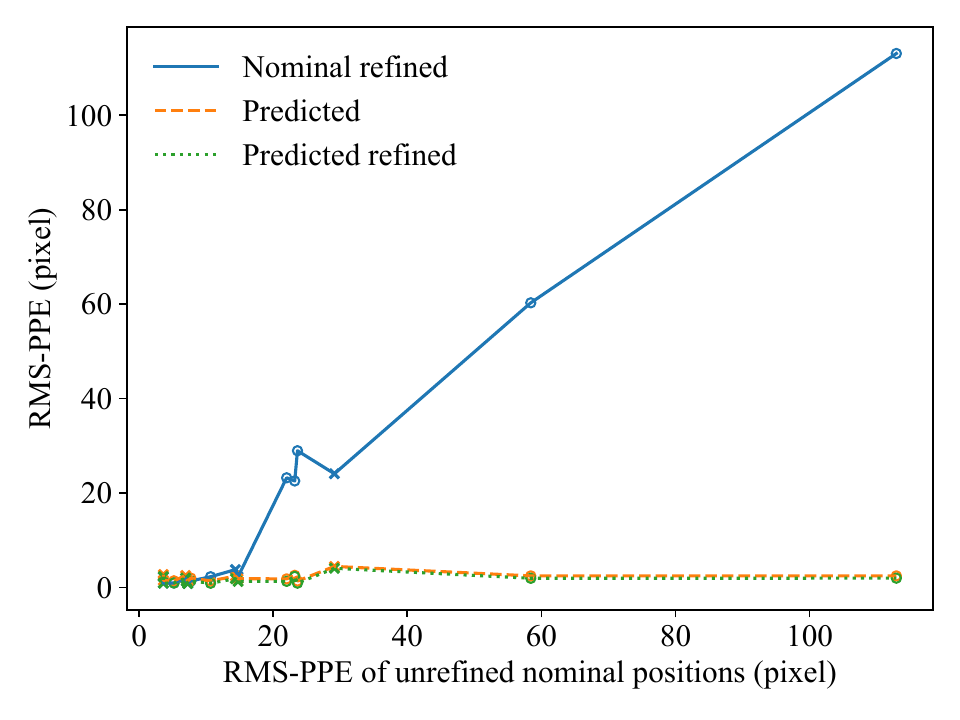}
    \caption{RMS-PPE of refined nominal, predicted, and
             refined predicted positions plotted against
             the RMS-PPE of unrefined nominal positions. Cases with
             accumulating errors are marked by ``$\circ$'', 
             while cases with independent errors are marked by ``$\times$''.}
    \label{fig:errors_against_nominal}
\end{figure}

To better visualize the performance of our method with different types and levels of
position errors, we plotted the RMS-PPE of refined nominal positions, predicted
positions, and refined predicted positions against the RMS-PPE of unrefined nominal
errors for all test cases in Fig.~\ref{fig:errors_against_nominal}.
Cases with accumulating and independent errors are respectively marked with 
``$\circ$'' and ``$\times$''. For the entire range of nominal RMS-PPEs, the accuracy
of predicted positions remains stable, while the RMS-PPE of refined nominal positions
starts to take off when the unrefined nominal RMS-PPE is around 15 pixels. Beyond this
inflection
point, our position prediction method is fundamentally advantageous over
optimization-based refinement alone. 

\begin{table}[]
    \centering
    \begin{tabular}{ccccc}
        \hline
        \multirow{2}{*}{\textbf{Label}} & \multicolumn{2}{c}{\textbf{Loss convergence}} & \multicolumn{2}{c}{\textbf{RMS-PPE convergence}} \\
        \cline{2-5}
         & Lower final loss? & Faster descent? & Lower final error? & Faster descent? \\
        \hline
        A1-1 & \checkmark & \checkmark & \checkmark & \checkmark \\ 
        A1-2 & \checkmark & \checkmark & \checkmark & \checkmark \\ 
        A1-3 & \checkmark & \checkmark & \checkmark & \checkmark \\ 
        A2-1 &  & \checkmark &  & \checkmark \\ 
        A2-2 & \checkmark & \checkmark & \checkmark & \checkmark \\ 
        A3-1 &  & \checkmark &  & \checkmark \\ 
        A3-2 &  & \checkmark & \checkmark & \checkmark \\ 
        A3-3 & \checkmark & \checkmark & \checkmark & \checkmark \\ 
        I1-1 &  &  &  & \\ 
        I1-2 &  & \checkmark &  & \checkmark \\ 
        I1-3 &  & \checkmark & \checkmark & \checkmark \\ 
        I1-4 & \checkmark & \checkmark & \checkmark & \checkmark \\ 
        I2-1 &  & \checkmark &  & \checkmark \\ 
        I2-2 &  & \checkmark & \checkmark & \checkmark \\ 
        I3-1 &  &  &  & \checkmark \\ 
        I3-2 &  & \checkmark &  & \checkmark \\ 
        \hline
    \end{tabular}
    \caption{The effects of initializing position refinement-enabled ptychographic reconstruction with predicted positions.}
    \label{tab:random_etch_loss_ppe}
\end{table}

More insights on how using predicted positions affects the convergence
behavior of ptychographic reconstruction loss and position refinement
during reconstruction can be obtained by looking at the
histories of the reconstruction
loss (in our case the mean-squared error between computed and measured far-field magnitudes)
and RMS-PPE for all datasets.
The plots are shown in Fig.~S3 in
Supplementary Materials, and the results are summarized in
Table~\ref{tab:random_etch_loss_ppe} in terms of whether initializing
reconstruction with predicted positions leads to lower final loss/RMS-PPE, or faster descent of loss/RMS-PPE at the 
early stage.
For the 3 cases with large accumulating
errors (A1-1 -- A1-3), initializing reconstruction with predicted positions is predominantly
more advantageous than using nominal positions, in that both the final reconstruction loss
and RMS-PPE are better than the latter scenario. This also agrees with the visual appearance
of the reconstructions in Fig.~\ref{fig:random_etch_recons_accumulating}(b) and (d). 
For smaller accumulating errors (A2-1 -- A3-3) and independent errors (I1-1 -- I3-2),
using predicted positions still lead to faster loss and RMS-PPE reduction in most cases, 
although it does not necessarily result in lower loss or RMS-PPE at convergence. 
I1-1 is the only exception where the use of predicted positions did not generate significant
benefit in either aspect, because the low magnitude and independent nature of the errors
pose a particularly easy case for optimization-based refinement.

A1-3 is the most notable case that clearly demonstrates the effectiveness of our position
prediction method. The RMS-PPE of the nominal positions is 112.93 pixels. After refinement using nominal positions,
the error in fact increased to 113.09 pixels, indicating that the gradients of the loss with regards
to positions are not reliable enough to push the positions to the right direction. 
The predicted positions
have a much lower RMS-PPE of 2.43 pixels, and after refinement it was further lowered to 1.95 pixels.

Although comparing the absolute probe positions could introduce bias due to the equivalence
of ptychographic positions to arbitrary global offsets, it could still offer intuitive 
insights in this particular case. 
The RMS of the deviation between mean-subtracted absolute nominal and true positions is 111.02 pixels (888.16 nm)
and the maximum is 231.64 pixels (1853.12 nm).
The predicted positions brought the RMS down to 3.19 pixels 
with a maximum point-to-point distance of 9.77 pixels, a level that is much easier for 
optimization-based algorithm to do further refinement. 

\begin{figure}
    \centering
    \includegraphics[width=0.95\textwidth]{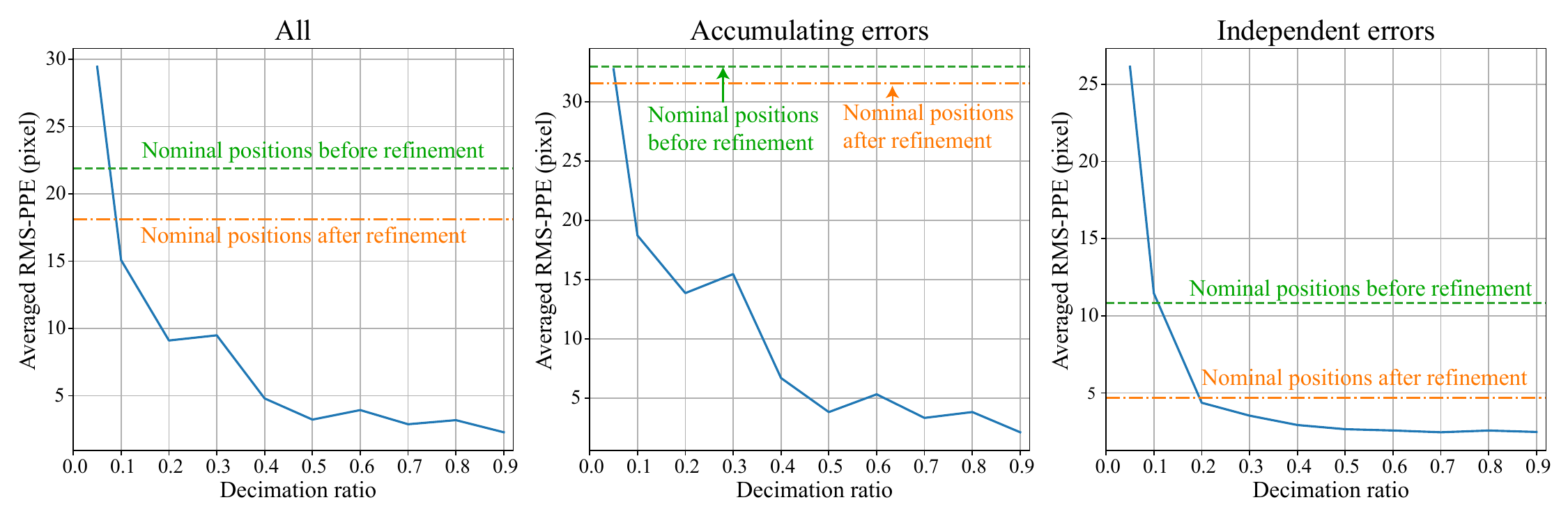}
    \caption{RMS-PPE of predicted positions (without refinement) averaged among 
    (a) all test cases, (b) cases with accumulating errors, and (c) cases with
    independent errors using images predicted by \ptychonn{}
    trained with reduced data sizes. The fraction of the training set size 
    in the full data size is
    denoted as the decimation ratio. The gray dashed lines respectively denotes the averaged
    RMS-PPE of nominal positions with and without optimization-based refinement.}
    \label{fig:decimation_test}
\end{figure}

While we trained \ptychonn{} with about 31.5 thousand diffraction patterns 
(excluding validation data), training data may be
scarce in reality. Therefore, stable performance under low-data scenario would be a desirable
characteristic for a deep learning-based approach. We conducted a data decimation
study, where we trained \ptychonn{} with various fractions of data of the full training set, where 
the fraction, hereafter referred to as the data decimation ratio,
ranges from 5\% to 90\%. We then conducted position prediction on all the datasets
for all decimation ratios.
As serial mode is more likely to be impacted by inaccurate phase images, we use
the nominal positions for nearest neighbor search in collective mode prediction. 
The results are plotted in Fig.~\ref{fig:decimation_test}, with
the subplots respectively showing
the RMS-PPEs of the predicted positions without refinement 
averaged across all cases (a), cases with accumulating errors (b), 
and cases with independent errors (c).
The plot shows a steady trend of RMS-PPE for both accumulating and independent errors 
when the decimation ratio is larger than 0.4,
where the averaged RMS-PPE is significantly lower than that of using nominal positions
after refinement. For accumulating errors, the error starts to rise when the decimation
ratio is below 0.4, but remains lower than that of nominal positions after
refinement until the decimation ratio is 5\%, at which point the predicted images
are no longer structurally consistent across adjacent tiles and features become heavily
distorted. For independent errors, the RMS-PPE of predicted positions remains lower
than refined nominal positions until the decimation ratio drops below 0.2. 

\subsection{Siemens star object}

\begin{figure}
    \centering
    \includegraphics[width=\textwidth]{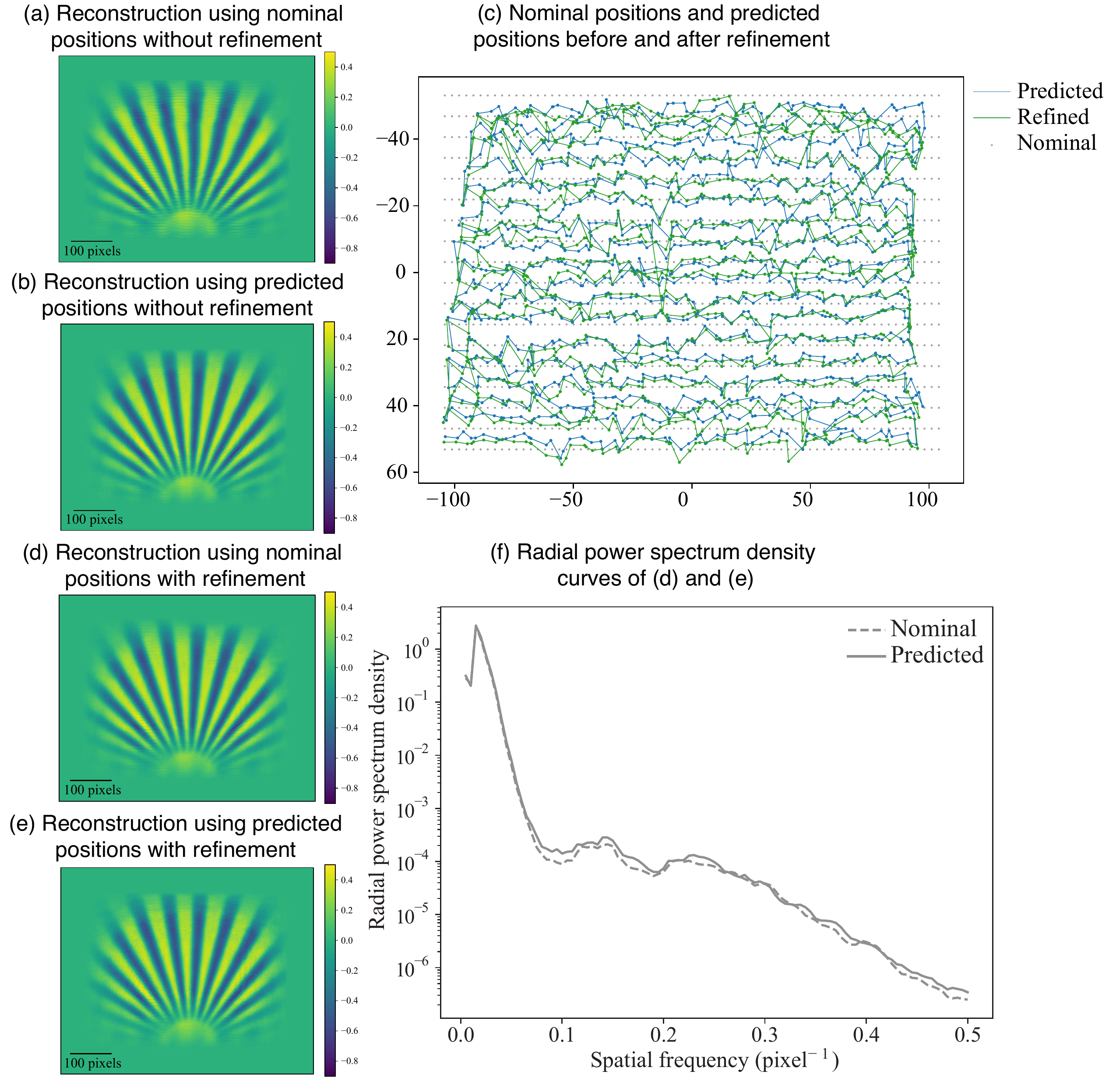}
    \caption{Results of probe position prediction and ptychographic reconstruction on the
             Siemens star dataset. (a) and (b) are the reconstructions using nominal and
             predicted positions without position refinement. After turning on position
             refinement during reconstruction, we obtained the refined positions that are
             plotted together with predicted positions in (c). (d) and (e) are the
             reconstructions with nominal and predicted positions after refinement. 
             (f) shows the radial PSD curves of the reconstructions with nominal and
             predicted positions. }
    \label{fig:siemens_star_results}
\end{figure}

The position errors in the Siemens star dataset occurred naturally during the experiment,
and is a representative case of combined accumulating errors and independent errors. While
ground truth positions are not available, we may visualize the errors by
reconstructing the data with nominal positions (a rectangular grid) and without position
refinement. The result is shown in Fig.~\ref{fig:siemens_star_results}(a), where one may see
not only localized jitters along the spokes of the Siemens star, but also a long-range
drift of positions revealed by the leftward skewing of the spokes near the top of the image. 
We ran our position prediction algorithm on the test dataset for one collective mode iteration
using nominal positions for nearest neighbor search. With the predicted positions, 
a reconstruction without position refinement is shown in Fig.~\ref{fig:siemens_star_results}(b),
where the distortions in (a) are fundamentally corrected. Another reconstruction run
with position refinement provides us with the refined positions, which are plotted together
with predicted positions in Fig.~\ref{fig:siemens_star_results}(c). The predicted 
and refined positions are reasonably close, indicating that the predicted positions are already
highly accurate. Notably, the predicted positions are able to capture the long-range skew
on the top of the scanned area, where the positions are rightward skewed,
reflecting the leftward distortion in Fig.~\ref{fig:siemens_star_results}(a). 
The reconstructions with nominal and predicted positions with further refinement are shown in
Fig.~\ref{fig:siemens_star_results}(d) and (e). While the images exhibit little visual
difference, we may quantitatively compare their spatial resolution by analyzing their
radial power specturm density (RPSD). The RPSD of an image $I$ is calculated as
\begin{equation}
    \mbox{RPSD} = \frac{\sum_{u, v} |\mathcal{F}[I](u, v)|^2 M_r(u, v)}{\sum_{u, v} M_r(u, v)}
    \label{eqn:rpsd}
\end{equation}
where $\mathcal{F}[I]$ is the Fourier transform of $I$ after normalized by the DC value,
and $M_r(u, v)$ is a mask whose values are 1 for reciprocal coordinates $u, v$ satisfying
$\sqrt{u^2 + v^2} = r$ and drops to 0 when $|\sqrt{u^2 + v^2} - r| \ge 0.5\delta_r$, with
$\delta_r$ being the step size of $r$. Essentially, $M_r(u, v)$ is a ring in the reciprocal
space with radius $r$ whose unit is reciprocal pixel; plotting the power spectrum density
averaged over the ring for a range of $r$ reveals the frequency component distribution of
a 2D signal. The RPSD plots for the reconstructions with nominal and predicted positions
are shown in Fig.~\ref{fig:siemens_star_results}(f). Both curves 
are almost identical in low-frequency regime below 0.08 pixel$^{-1}$, but the reconstruction
with predicted positions have stronger Fourier component in mid-to-high frequency regime.
Since the images have very low noise level based on both their visual appearance and the lack
of high-frequency plateaus in the RPSD plot, our observation indicates that the reconstruction
with predicted positions contains stronger high-frequency signal, and is thus sharper
than the reconstruction with nominal positions. 

\section{Discussion}

\subsection{Robustness}

Single-shot phase prediction is the first and most important stage in our method. In Section
\ref{sec:random_etch_sample}, we have shown the result of a data decimation test, which indicates
that the algorithm's RMS-PPE remains overall stable when trained with about 40\%, or $1.3\times 10^4$
diffraction patterns. Considering that the area with high sample-to-sample variation
(which mostly coincides with the far-field diffraction pattern that would be formed by an
unscattered probe) consists of about 1000 pixels, we estimate the total number of data points
to be on the order of $10^7$. The phase-only model's parameter size is about $7.4\times 10^5$,
so the over-determination ratio is 10 -- 20. This is close to the lower bound given by
the widely adopted rule-of-thumb that the data points needed to train a generalizable model
should be at least 10 times of the Vapnik–Chervonenkis 
dimension of the model, which roughly equals to
the number of parameters 
\cite{baum_neuralcomputation_1989,abumostafa_neuralcomputation_1995,
alwosheel_j_choice_modeling_2018}. The Siemens star experiment had an even smaller training set
of 2450 diffraction patterns or about $2\times 10^6$ high-variance pixels, 
yet we were still able to obtain predicted positions that
are reasonably accurate themselves and lead to sharper images after reconstruction and
refinement. Our adaption of batch normalization and mean-subtraction feature engineering
to the vanilla \ptychonn{} architecture
have contributed to the low-data stability and generalizability of the model. 
Meanwhile, self-supervised phase retrieval networks
\cite{yao_npjcompmat_2022, hoidn_scirep_2023} have been reported recently. These networks
incorporate physical models and are trained using only the diffraction patterns without needing
real-space reconstruction as labels. This technique is promising in addressing the concern of
training data scarcity. 

Image registration is another important part of our algorithm. If the nearest neighbors for
each scan point can be well determined either through serial mode registration or from
nominal positions, the collective mode registration is robust to sporadic registration errors
since multiple neighbors are registered for each scan point, and the probe positions are
solved as the least-squares solution of a linear problem involving all the registered pairs. 
However, an overly large number of nearest neighbors may actually reduce position prediction
accuracy, because more largely spaced pairs that are 
harder to register
would be involved. To overcome this problem, we reject registration results if their errors
between the images after applying the offset exceed a set threshold. 
Yet, over-stringent registration screening
is likely to reject all nearest neighbors for a certain point, making it unconstrained. 
The error threshold and the number of nearest neighbors need to be chosen carefully in order for optimal performance. 

\subsection{Computation cost}

With regards to computation cost, an error map registration on a $128\times 128$-sized
image pair with a search range of $\pm 20$ pixels takes about 0.6 s on CPU. The faster SIFT
algorithm takes about 0.08 s. In reality, the average registration time might be longer:
for error map, the algorithm needs to search in a larger range if the lowest error found
in the current search range is still too high; when hybrid registration is used, the algorithm
may need to run both error map and SIFT sequentially if the error map pass does not find
an offset with sufficiently small error. Once the pairwise registration finishes, solving
the linear system can be done in seconds for a moderate-size dataset
using direct inversion (1 -- 2 second for random etched object
datasets with 961 diffraction patterns each). This walltime can stay controllable for larger
scans by using iterative matrix inversion algorithms, GPU acceleration, and sparse linear
algebraic approaches. On the other hand, the registration part of the algorithm
scales linearly with the number of scan points and nearest neighbors. The pairwise registration
tasks are also independent across different pairs of points, opening the possibility to accelerate the process
through parallelization. 

\subsection{Comparison with multi-level refinement}

While our method has demonstrated advantages dealing with large accumulating errors, it should
be recognized that optimization-based algorithms may also get improved performance on such
errors by adopting a multi-level strategy, where one runs reconstruction with position
refinement with large real-space pixel size, so that the error distances shrink to fewer
pixels; after that, one may progressively repeat the 
refinement with smaller pixel sizes for better precision. 
One limitation of this approach, however, is that the pixel size cannot be increased indefinitely
especially for certain types of probe functions. The real space pixel size of ptychography
is determined by $\delta_o = \lambda z / L \delta_r$, where $\lambda$ is the wavelength,
$z$ is the sample-to-detector distance, $L$ is the side length of the diffraction pattern
in pixels, and $\delta_r$ is the reciprocal pixel size. To increase $\delta_o$, one needs
to either reduce $L$ which means cropping the diffraction patterns, or reduce $\delta_r$
which means magnifying the diffraction patterns. In a zone plate-based setup where a central
beam stop is often used to block directly transmitted beams, the diffraction patterns have
a donut-like shape where the central region hardly contains any usable signal. Increasing the
real-space pixel size by a factor more than 3 or 4 would likely require the high-signal regions
to be cropped or pushed out of the image support, leaving only the central signal-absent region.
Additionally, for objects that lacks long-ranged variation and contain only fine features 
whose length scale is close to the
original real-space pixel size, radically enlarging the pixel size could remove the
high-frequency signals that make diffraction patterns distinguishable, causing position
correction to fail. 

\subsection{Challenges}

Our method so far is still faced with several challenges. First, during our experiment, we
found that the training and prediction of \ptychonn{} are more difficult on weakly
scattering objects, such as soft biological tissues. Diffraction patterns in such cases have
very little variation when examined visually, and we found the trained model hard to generalize
on the test set even with our mean subtraction approach for data transformation. 
Prediction quality on strong phase objects are better,
but they are not absolutely precise as we have observed slight deformation
between matching features in the predicted images of adjacent scan points,
especially if \ptychonn{} is asked to predicted the object function in an area that is much
larger than the region illuminated by the probe.
This issue might intensify for samples that are structurally complex 
(\emph{e.g.}, biological samples with a wide distribution
of feature scale, morphology, and material), where more training data and larger model capacity
might be needed to yield satisfactory performance.
Improving the prediction accuracy and consistency is among our future works. 
Addressing
this issue might require more sophisticated feature engineering measures such as designing
a look-up table to reduce the dynamic range of the diffraction patterns without compromising
the signals. Meanwhile, conditional generative models, especially a diffusion process guided
by physics models \cite{chung_nips_2022}, point to a potential direction for \ptychonn{}'s 
future evolution. The fact that diffusion models explicitly sample from a learned distribution,
along with the physics-based conditioning, poses strong constraints that might allow \ptychonn{}
to make high-quality predictions even for weak-phase or structurally complicated objects. 

The second challenge is also surrounding \ptychonn{}, in that one may need to re-train, or fine tune
the model to predict on data collected with very different probe functions, or from samples
that have little similarity in feature size, shape, or histogram distribution to the training data.
This challenge comes from the fact that \ptychonn{} implicitly learns the probe and the sample
image's distribution; both function as the prior knowledge for \ptychonn{} to work under the
information deficiency due to the absence of overlapping diffraction patterns. Improving the 
generalizability of \ptychonn{} might need one to supply prior knowledge also through known
physical laws instead of through data alone. Physics-informed self-supervision approaches
\cite{yao_npjcompmat_2022, hoidn_scirep_2023} could provide a viable path to this objective. 

A third challenge of our method is regarding the performance on samples that are spatially
sparse or dominated by one-dimensional features. Image registration is unable to find a
unique offset solution for images that are blank or contain only 1D features. We have
observed this issue when predicting the positions for the Siemens star object in regions
far away from the center, where a predicted tile may contain only one edge or at most 2 edges that
are almost parallel. As a remedy, we can detect such cases and use the offsets from
nominal positions in lieu of registration results if they are available, and our 
collective mode algorithm prevents the spread of errors from using nominal offsets
by calculating a least-squares solution of an overall well-constrained problem.

Lastly, our method may still require manual tuning of key parameters like the number
of nearest neighbors and the error tolerance for image registration. Nevertheless, the needs of parameter
tuning can also be relaxed with an improved \ptychonn{} that predicts sharp, accurate, and
consistent images, which greatly enhance the success rate of image registration.

\section{Conclusion}

We have developed a method for predicting ptychographic probe positions directly from diffraction
patterns using a single-shot phase retrieval neural network, \ptychonn{}. With a well trained
\ptychonn{} model, our method can achieve satisfactory prediction accuracy for data with large
and accumulating position errors beyond the capability of optimization-based position correction
algorithms. Our method has been shown to reduce accumulating errors with an average
deviation distance of $10^2$ pixels down to a few pixels. One may then use the predicted positions
to initialize optimization-based position refinement algorithms during ptychographic reconstruction
to reach better positional accuracy. For small or non-accumulating errors that could be corrected
by optimization-based algorithms eventually, initializing with our predicted positions can
still accelerate loss convergence and position error reduction. 
With future efforts invested in improving \ptychonn{}'s generalizability and prediction quality
on weak-phase objects, 
we expect our method to
be helpful in addressing a major challenge for ptychographic experiment stations 
without sophisticated
position tracking instruments such as high-precision interferometer systems. 

\section*{Funding}
U.S. Department of Energy (DOE), Office of Science, Basic Energy Sciences (DE-AC02-06CH11357).

\section*{Acknowledgments}
Work performed at the Center for Nanoscale Materials and Advanced Photon Source, both U.S. Department of Energy Office of Science User Facilities, was supported by the U.S. DOE, Office of Basic Energy Sciences, under Contract No. DE-AC02-06CH11357.

\section*{Disclosures}
The authors declare no conflicts of interest.

\section*{Program and data availability}
The codes of the position prediction algorithm, including those that defines and trains
\ptychonn{}, are openly available on our GitHub repository 
\url{https://github.com/mdw771/probe_position_correction_w_ptychonn}. 
Data for demonstration are also available on the repository. The rest of the data presented
in this manuscript are available upon request from the corresponding authors. 

\section*{Supplemental document}
See Supplement 1 for supporting content. 

\bibliographystyle{unsrt}
\bibliography{mybib,mybib1,mybib2}

\end{document}


\maketitle

\begin{abstract}
The supplemental information for the manuscript
\emph{Predicting ptychography probe positions using single-shot phase retrieval neural network} is presented in this document.
\end{abstract}

\section{Cross-validation of ptychographic reconstructions with\\PtychoShelves}

\begin{figure}
    \centering
    \includegraphics[width=\textwidth]{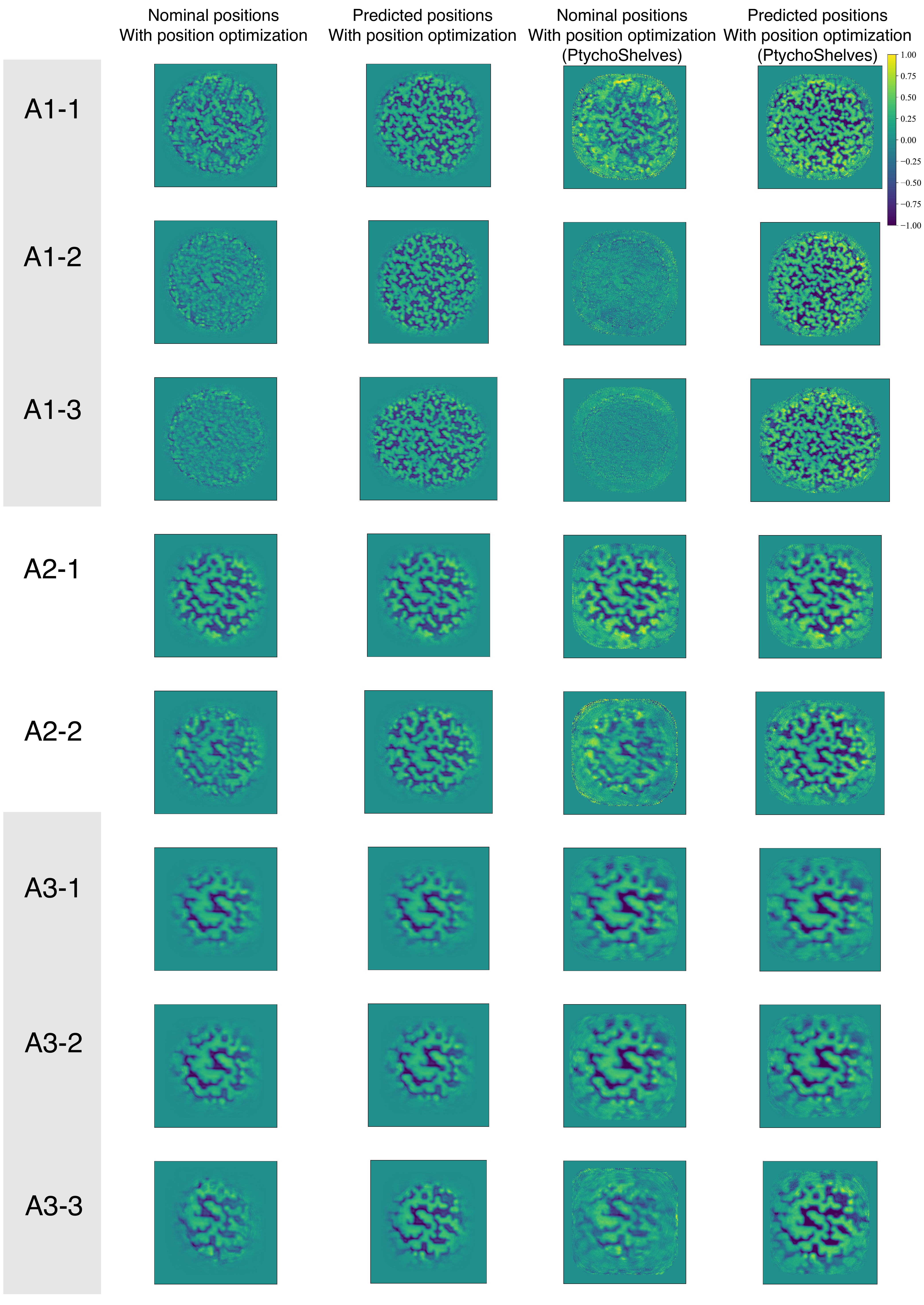}
    \caption{Comparison of \tike{} and \ptychoshelves{} reconstructions for datasets with
             accumulating errors. }
    \label{fig:random_etch_ptychoshelves_accumulating}
\end{figure}

\begin{figure}
    \centering
    \includegraphics[width=\textwidth]{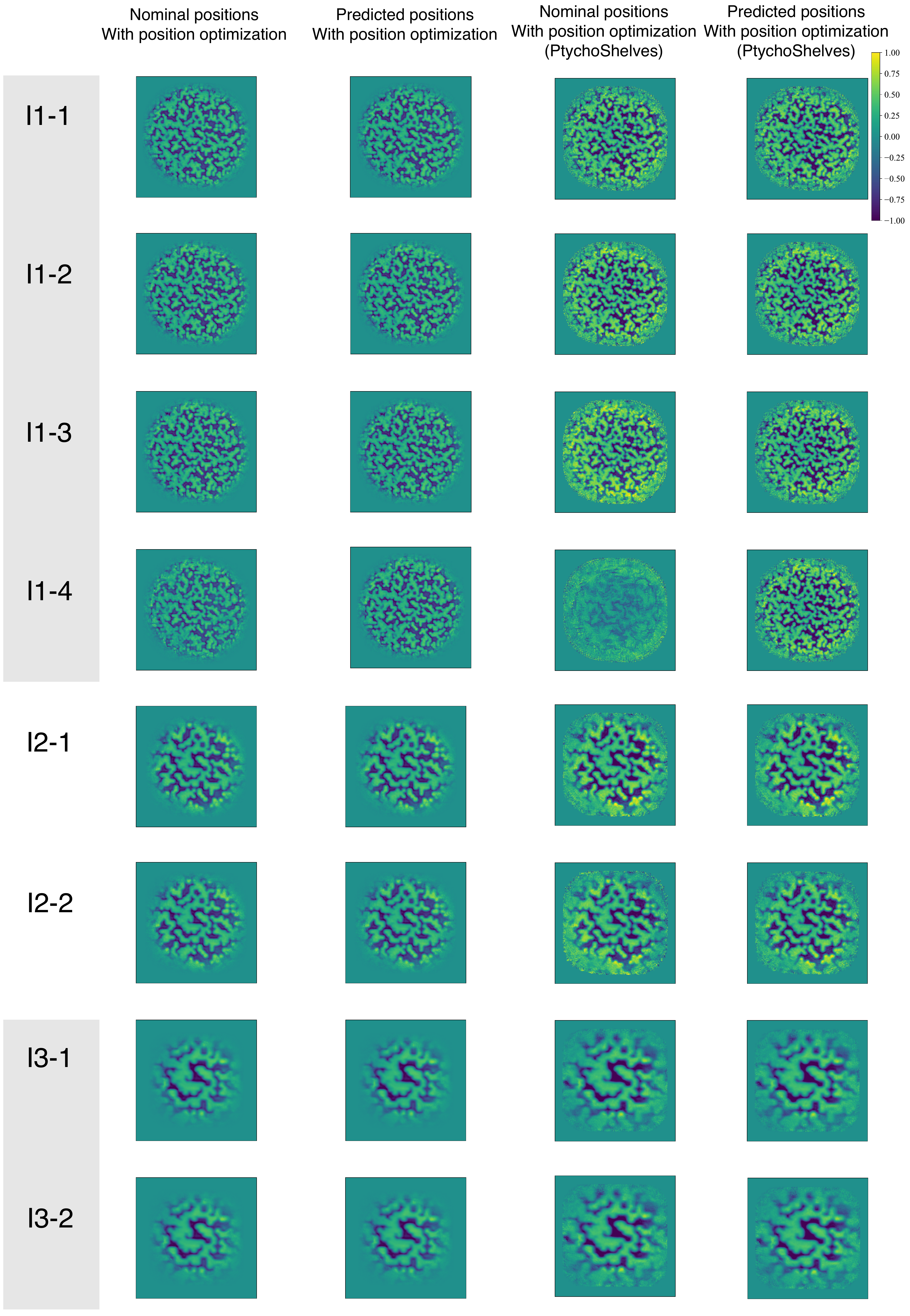}
    \caption{Comparison of \tike{} and \ptychoshelves{} reconstructions for datasets with
             independent errors. }
    \label{fig:random_etch_ptychoshelves_indpendent}
\end{figure}

The ptychographic reconstructions shown in the main manuscript are conducted using
\tike{}, a toolbox for ptychographic tomography reconstruction being developed at
the Advanced Photon Source. To validate the reconstruction results of \tike{}, we
also reconstructed the data with probe position refinement using 
a published ptychographic reconstruction software \ptychoshelves{}
\cite{wakonig_jac_2020}, and compare the results to guarantee their consistency. 
Fig.~\ref{fig:random_etch_ptychoshelves_accumulating}
and \ref{fig:random_etch_ptychoshelves_indpendent} shows the reconstructions of all data presented in
Fig.~3 and 5 of the main manuscript with \tike{} and \ptychoshelves{} side-by-side. 
Both software used rPIE \cite{maiden_optica_2017} for phase retrieval;
for probe position correction, \tike{} uses the Adam optimizer \cite{kingma_arxiv_2014} for 
position update, while \ptychoshelves{} uses a linearized least-squares method documented
in \cite{odstrcil_opex_2018}. As shown in the figures, \ptychoshelves{} also failed 
at the cases where \tike{} did not give a structure with
recognizable features, and vice versa. The only exception is case I1-4, where
\tike{} was able to deliver a mediocre reconstruction
with nominal positions
although it is still blurry due to uncorrected position error, while \ptychoshelves{}
did not manage to reach a reconstruction with the structures expected. 

The purpose of this comparison is not to prove the exact equivalence between \tike{}
and \ptychoshelves{} but to demonstrate that the advantage of using predicted positions
versus using nominal positions is consistent across reconstruction tools and algorithms. 
From this point, the comparison agrees with our analysis in the main manuscript. 

\section{Reconstruction loss and RMS-PPE history}

\begin{figure}
    \centering
    \includegraphics[width=\textwidth]{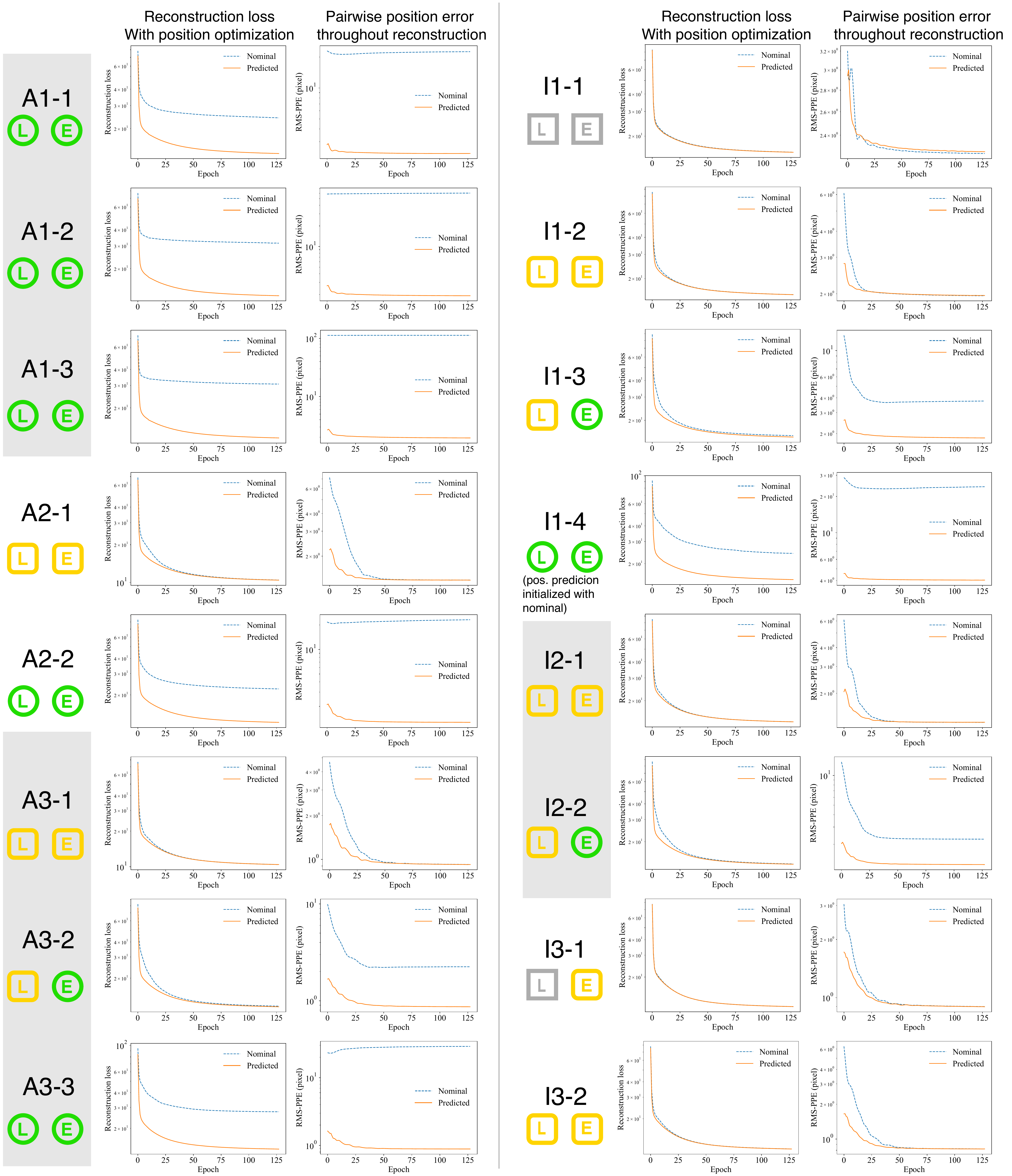}
    \caption{The reconstruction loss and RMS-PPE history of all datasets from the
             random etched object. 
             The leftmost column shows the dataset indices and a summary of how
             using predicted positions as the initial guess for position correction-enabled
             ptychography reconstruction affects the converge of reconstruction loss (symbolized by ``L'') and
             RMS of pairwise position errors (RMS-PPE; symbolized by ``E''). 
             A green, circled symbol 
             means using predicted
             positions leads to lower values of that quantity (loss or RMS-PPE) when the reconstruction
             finishes, and an orange, rounded corner-boxed 
             one means the final values are similar
             but using predicted positions leads to faster reduction. A gray, square-boxed symbol indicates no
             significant difference is found. }
    \label{fig:random_etch_loss_ppe}
\end{figure}

Fig.~\ref{fig:random_etch_loss_ppe} shows the plots of reconstruction loss
and RMS-PPE history of all test cases from the random etched object. The symbolic indicators ``L'' and ``E'',
whose meanings are explained in the figure's caption, 
offer an overview of how using predicted positions as the initial position guess affects the 
reconstruction and position refinement result in \tike.
The results are  also summarized in Table 
2
in the main text.

\bibliographystyle{unsrt}
\bibliography{mybib,mybib1,mybib2}